\documentclass{article}
\usepackage{amsmath}
\usepackage{graphicx}
\usepackage{hyperref}
\usepackage{authblk}
\allowdisplaybreaks
\begin{document}
\title{Observables from a perturbative, accelerating solution of relativistic hydrodynamics%
\thanks{Presented at XIII Workshop on Particle Correlations and Femtoscopy, 22-26 May 2018, The Henryk Niewodnicza\'nski Institute of Nuclear Physics PAN, Krak\'ow, Poland}%
}
\author[1]{ B\'alint Kurgyis} 
\author[1]{M\'at\'e Csan\'ad}
\affil[1]{E\"otv\"os Lor\'and University, Hungary\\H-1117 Budapest, P\'azm\'any P. s. 1/A}

\maketitle
\begin{abstract}
The discovery of the almost perfect fluid like nature of the strongly interacting quark-gluon plasma was one of the most important discoveries of heavy-ion physics in recent decades. The experimental results are well described by hydrodynamical models. Most of these models are numerical simulations, however the analytic solutions are also important in understanding the time evolution of the quark-gluon plasma created in the heavy-ion collisions. Here we present a perturbative, accelerating solution on top of a known solution, the relativistic Hubble flow. We describe the perturbative class of solutions, and calculate a few observables for a selected solution.  
\end{abstract}
%\PACS{25.75.-q, 24.10.Nz}
  
\section{Introduction}
The quark-gluon plasma was discovered in heavy-ion collisions at RHIC \cite{Adcox:2004mh, Adams:2005dq} and it was also created at LHC \cite{Aamodt:2010pa,Aamodt:2010jd}. Relativistic hydrodynamics can be utilized to describe the properties of the medium  from the initial thermalization ($\sim 1\mathrm{~fm/}c$) until the hadronization  ($\sim 10\mathrm{~fm/}c$), when hadrons are created and freeze-out takes place.
To obtain the equations of hydrodynamics one has to consider the energy and momentum conservation, which can be formalized this way:
\begin{equation}\label{eq:energy-momentum}
\partial_\mu T^{\mu\nu}=0,
\end{equation}
where $T^{\mu\nu}$ is the energy-momentum tensor. In the case of perfect fluid hydrodynamics it is the following:
\begin{equation}
T^{\mu\nu}=(\varepsilon+p)u^\mu u^\nu-pg^{\mu\nu},
\end{equation}
with $u^\mu$ being the flow field (where the $u_\mu u^\mu=1$ constraint applies), $\varepsilon$ the energy density and $p$ the pressure. This tensor equation can be projected into a time-like and a space-like equation to make the calculations easier. By performing the projections we get the Euler and the energy equations:
\begin{align}\label{eq:euler}
(\varepsilon+p)u^\mu\partial_\mu u^\nu&=(g^{\mu\nu}-u^\mu u^\nu)\partial_\mu p,\\\label{eq:energy}
u^\mu\partial_\mu \varepsilon+(\varepsilon+p)\partial_\mu u^\mu&=0.
\end{align} 
In addition to conserved energy and momentum, we assume that the system has some conserved charge density $n$:
\begin{equation}\label{eq:conservation}
\partial_\mu(nu^\mu)=0.
\end{equation} 
If there are no conserved charges (at zero chemical potential) we can derive a similar conservation equation for the entropy density $\sigma$.
To get a full set of equations one has to introduce thermodynamical relations, for this we take the Equation of State (EoS) in the form of $\varepsilon=\kappa p$, where $\kappa$ is a constant during our calculations, but we have to note that it could be a temperature dependent $\kappa(T)$ \cite{Csanad:2012hr}. These equations should be solved in order to get the time evolution of a relativistic fluid. We can do this either by numerical simulations or looking for analytic solutions. The latter is a rather challenging task, but also it gives us a better understanding about the connection between the initial and final state of the matter. Now we are focusing on analytic solutions.
\section{The known solution: Hubble flow}
For the above equations several different sets of solutions exist. Historically the first ones were 1+1D solutions: the Landau-Khalatnikov \cite{Landau:1953gs,Khalatnikov:1954aa} and the Hwa-Bjorken \cite{Hwa:1974gn,Bjorken:1982qr} solutions. Inspired by the QGP and its fluid nature new solutions were found. One important example is the first relativistic 3+1D solution with realistic 
geometry, the Hubble flow solution of ref.~\cite{Csorgo:2003ry}, describing a self-similar expansion.
The geometry of the solution is characterized by the scaling variable $S$ and an arbitrary function $\mathcal{N}(S)$ appearing in the density distribution. The scaling variable has vanishing comoving derivative ($u_\mu\partial^\mu S=0$). The importance of this solution in heavy-ion physics is that it describes well the measured experimental data of hadrons and even photons \cite{Csanad:2011jq,Csanad:2009wc}. This solution describes an acceleration-less ($u_\mu\partial^\mu u^\nu=0$) Hubble flow. We would like to describe a system ``similar'' to the original Hubble flow, but with small acceleration and pressure gradient. For this we study the perturbations on top of this known solution.
\section{The perturbations of the Hubble flow}
To get the perturbations on top of the Hubble flow, here we follow the same method as in ref.~\cite{Kurgyis:2017zxg} and the idea is similar to that of ref.~\cite{Shi:2014kta}. To begin with, we introduce perturbations of hydrodynamical fields ($u^\mu\rightarrow u^\mu+\delta u^\mu,p\rightarrow p+\delta p,\dots$) to be substituted to the general equations of hydrodynamics. Then we subtract the zeroth order equations, since they describe an already known solution, and we neglect the second or higher order terms of perturbations. This way we end up with the general first order equations, as given in ref.~\cite{Kurgyis:2017zxg}, and after this we have to substitute the fields of Hubble flow.
In addition we can derive a constraint for the four-velocity perturbation by considering the normalization of the four velocity: $u_\mu\delta u^\mu=0$.
We found a set of general solutions in the following form:
\begin{align} \label{du01}
\delta u^\mu&=\delta \cdot F(\tau) g(x_\mu)\chi (S)\partial^\mu S,\\ \label{dp01}
\delta p&=\delta\cdot p_0\left(\frac{\tau_0}{\tau}\right)^{3+\frac{3}{\kappa}}\pi (S),\\ \label{dn01}
\delta n&=\delta \cdot n_0\left(\frac{\tau_0}{\tau}\right)^3 h(x_\mu)\nu (S).
\end{align}
Here $\delta$ is the perturbation scale and $g$, $h$, $F$, $\chi$, $\pi$, $\nu$ are arbitrary functions of $x^\mu$, $\tau$ and $S$, respectively, under the following restriction equations:
\begin{align}\label{chi:1}
\frac{\chi'(S)}{\chi(S)}&=-\frac{\partial_\mu\partial^\mu S}{\partial_\mu S\partial^\mu S}-\frac{\partial_\mu S \partial^\mu \ln g(x_\mu)}{\partial_\mu S\partial^\mu S},\\ \label{pi:1}
\frac{\pi'(S)}{\chi(S)}&=(\kappa+1)\left[F(\tau)\left(u^\mu\partial_\mu g-\frac{3g(x_\mu)}{\kappa\tau}\right)+F'(\tau)g(x_\mu)\right],\\ \label{nu:1}
\frac{\nu (S)}{\chi(S)\mathcal{N}'(S)} &=-\frac{F(\tau)g(x_\mu) \partial_\mu S\partial^\mu S}{u^\mu\partial_\mu h(x_\mu)}.
\end{align}
It is important to observe, the terms on the left of each restriction equation are functions of solely $S$, therefore this should apply to the right hand sides as well. Therefore we can not find a solution for these equations with any scaling variable with vanishing comoving derivative, but only with appropriately chosen $F$, $h$, $g$ and an $S$ accordingly.
\section{A concrete solution}\label{sect:s:tr}
\begin{figure}[]
\centerline{%
\includegraphics[trim=8 10 8 150, clip,width=12.5cm]{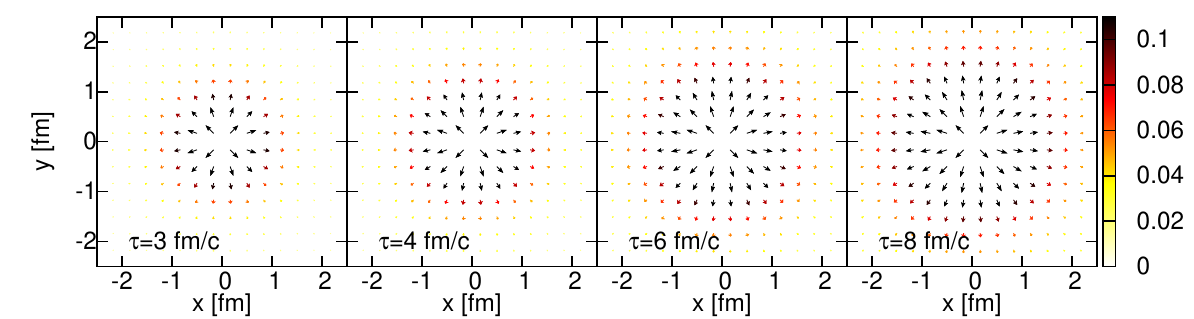}}
\caption{The proper time evolution of the four-velocity perturbation in the $x-y$ plane.}
\label{fig:du2d}
\end{figure}

Here we shortly discuss a simple, concrete solution that is investigated in more details in ref.~\cite{Kurgyis:2017zxg}. For the scaling variable we chose the spherically symmetric $S=t/r$, the $m=-1$ case of ref.~\cite{Kurgyis:2017zxg} section 4. The $F$, $g$, $h$ functions are the following:
\begin{align}
F(\tau)&=\tau+c\tau_0\left(\frac{\tau}{\tau_0}\right)^{\frac{3}{\kappa}},\\
g(x_\mu)&=1,\\
h(x_\mu)&=\begin{cases}
              \ln\left(\frac{\tau}{\tau_0}\right)+ c\frac{\kappa}{3-\kappa}\left(\frac{\tau}{\tau_0}\right)^{\frac{3}{\kappa}-1}&\textnormal{ if } \kappa\neq 3\\
              (1+c)\ln\left(\frac{\tau}{\tau_0}\right)&\textnormal{ if } \kappa=3
		  \end{cases}
\end{align}
where $c$ is an arbitrary parameter of the solution. For the charge density we assume a Gaussian profile: $\mathcal{N}(S)=\exp{(-{r^2}/{t^2})}=\exp{(-S^{-2})}$. To study this solution we used the original Hubble flow model parameters of refs.~\cite{Csanad:2011jq,Csanad:2009wc}, where hadron and photon spectra were fitted with observables calculated from the original Hubble flow. An example plot for the proper time evolution of the four velocity perturbation is shown on fig.~\ref{fig:du2d}. We found that for small radial distances and large values of $\delta$ the perturbations give a large contribution to the system, comparable with the original solution. This sets a limit to the applicability for this given solution, however this is not necessarily a general property of this class of perturbations.
\section{Observables}
\begin{figure}[]
\centerline{%
\includegraphics[width=6.25cm]{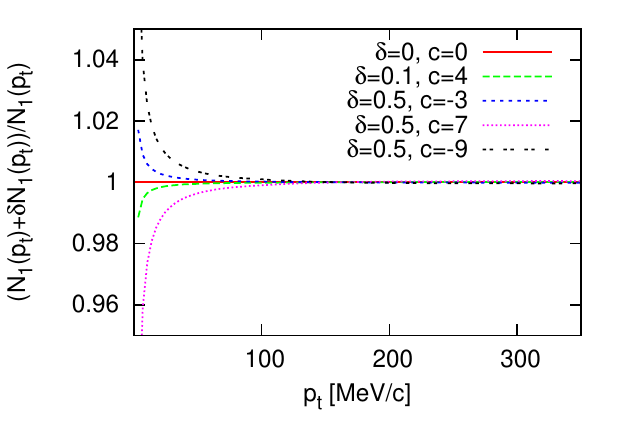}
\includegraphics[width=6.25cm]{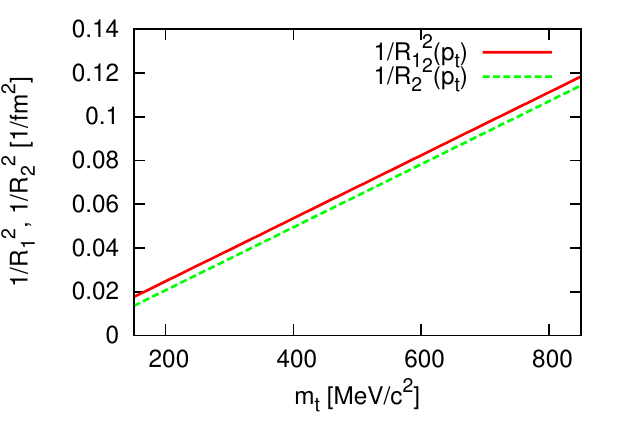}
}
\caption{The ratio of the original and the perturbed single-particle momentum distribution calculated for the particular solution from section~\ref{sect:s:tr} is in the left plot. In the right plot the transverse mass scaling of the HBT radii is shown.}
\label{fig:observables}
\end{figure}

Let us now focus on observables that can be calculated from the investigated hydrodynamical solutions. The source function is given by a Maxwell--J\"uttner distribution similarly as in ref.~\cite{Csanad:2009wc}:
\begin{equation}
S(x,p)=N n \exp\left(-\frac{p_\mu u^\mu}{T}\right) H(\tau)p_\mu \mathrm{d}^3\Sigma^\mu(x^\mu) \mathrm{d}\tau,
\end{equation}
where $N$ is a normalization factor, $T$ is the temperature ($T=p/n$). The $p_\mu \mathrm{d}^3\Sigma^\mu(x^\mu)$ term is the Cooper-Frye factor, and assuming that the freeze-out happens at a constant $\tau_0$ proper time the $H(\tau)$ becomes a Dirac-delta: $H(\tau)=\delta(\tau-\tau_0)$. The perturbed source function can be calculated by substituting the perturbed fields and neglecting the second or higher order terms. The perturbed source function is the following:
\begin{align}
S(x,p)&=N n \exp\left(-\frac{p_\mu u^\mu }{T}\right)\delta(\tau-\tau_0)\frac{p_\mu u^\mu }{u^0}(1+\Delta)\mathrm{d}\tau \mathrm{d}^3x,\\
\Delta&=\left[\frac{\delta u^0}{u^0}+\frac{p_\mu \delta u^\mu}{p_\nu u^\nu}-\frac{p_\mu \delta u^\mu}{T}+\frac{p_\mu u^\mu \delta T}{T^2}+\frac{\delta n}{n} \right].
\end{align}
We can calculate the single-particle momentum distribution by integrating the source function over the space-time coordinates:
\begin{equation}
N_1(p)=\int S(x,p) \mathrm{d}\tau \mathrm{d}^3x.
\end{equation}
The left plot of fig.~\ref{fig:observables} shows the ratio of the original and the perturbed transverse momentum distributions for different values of $c$ and $\delta$. The same concrete solution was used as in section~\ref{sect:s:tr}, and in ref.~\cite{Kurgyis:2017zxg} with model parameters from refs.~\cite{Csanad:2011jq,Csanad:2009wc}. For the calculation a Gaussian saddlepoint approximation was used. The perturbations are small except at $p_\mathrm{T}\sim0$ MeV/$c$. We also calculated the one dimensional HBT radius (femtoscopic homogeneity length) \cite{HanburyBrown:1956bqd, Goldhaber:1960sf} for this particular solution. The right plot of fig.~\ref{fig:observables} shows the $R_{\mathrm{HBT}}^2\propto 1/\sqrt{m_\mathrm{t}}$ transverse mass scaling of the HBT radii. The perturbed source is a two component Gaussian, therefore we have two radii, $R_1$ is equal to the original radius, and the perturbed radius is some average of $R_2$ and $R_1$. The perturbations cause only small deviation from the original values of the HBT radii. 
\section{Summary} 
We found a new class of accelerating, perturbative solutions on top of the relativistic Hubble flow. In the investigated particular case the hydrodynamical fields gave large contributions for small radial distances, which is a limiting factor to the applicability of that concrete solution. However, by calculating some observables, we saw that in the case of experimentally measurable quantities the perturbations cause only small deviations from the original observables.
%uncomment the following lines to place a figure
%\begin{figure}[]
%\centerline{%
%\includegraphics[width=12.5cm]{figure}}
%\caption{An example plot .}
%\label{fig:}
%\end{figure}

\section*{Acknowledgements}
The authors express gratitude for the support of Hungarian NKIFH grant No. FK-123842. B.~Kurgyis was supported by the \'UNKP-18-1-I-ELTE-320 New National Excellence Program of the Hungarian Ministry of Human Capacities. M.~Csan\'ad was supported by the \'UNKP-18-4 New National Excellence Program of the Hungarian Ministry of Human Capacities.

%\bibliography{ref}
\bibliographystyle{unsrt}

\end{document}